\documentstyle[12pt]{article}

        \input epsf
        \input amssym.def
        \input amssym.tex

\newcommand{\beq}{\begin{equation}}
\newcommand{\eeq}{\end{equation}}
\newcommand{\beqa}{\begin{eqnarray}}
\newcommand{\eeqa}{\end{eqnarray}}
\newcommand{\sect}[1]{\setcounter{equation}{0}\section{#1}}
\newcommand{\rf}[1]{(\ref{#1})}

\newcommand{\eps}{\epsilon}
\newcommand{\cch}{\cos\chi}
\newcommand{\csq}{\cos^2\!\chi}
\newcommand{\re}{{\rm Re}}
\newcommand{\im}{{\rm Im}}

\renewcommand{\theequation}{\arabic{section}.\arabic{equation}}

\begin{document}

\title{Pair Creation of Black Holes During Inflation}
\author{{\sc Raphael Bousso}\thanks{\it R.Bousso@damtp.cam.ac.uk} \ and
        {\sc Stephen W. Hawking}\thanks{\it S.W.Hawking@damtp.cam.ac.uk}
      \\[1 ex] {\it Department of Applied Mathematics and}
      \\ {\it Theoretical Physics}
      \\ {\it University of Cambridge}
      \\ {\it Silver Street, Cambridge CB3 9EW}
       }
\date{DAMTP/R-96/33}

\maketitle

\begin{abstract}

Black holes came into existence together with the
universe through the quantum process of pair creation in the
inflationary era. We present the instantons responsible for this
process and calculate the pair creation rate from the no boundary
proposal for the wave function of the universe. We find that this
proposal leads to physically sensible results, which fit in with other
descriptions of pair creation, while the tunnelling proposal makes
unphysical predictions.  

We then describe how the pair created black holes evolve during
inflation. In the classical solution, they grow with the horizon scale
during the slow roll--down of the inflaton field; this is shown to
correspond to the flux of field energy across the horizon according to
the First Law of black hole mechanics. When quantum effects are taken
into account, however, it is found that most black holes evaporate
before the end of inflation. Finally, we consider the pair creation of
magnetically charged black holes, which cannot evaporate. In standard
Einstein--Maxwell theory we find that their number in the presently
observable universe is exponentially small. We speculate how this
conclusion may change if dilatonic theories are applied.

\end{abstract}

\pagebreak

\sect{Introduction}

It is generally assumed that the universe began with a period of
exponential expansion called inflation. This era is characterised by
the presence of an effective cosmological constant $\Lambda_{\rm eff}$
due to the vacuum energy of a scalar field $\phi$. In generic models
of chaotic inflation~\cite{Lin83,Haw84}, the effective cosmological
constant typically starts out large and then decreases slowly until
inflation ends when $\Lambda_{\rm eff} \approx 0$. Correspondingly,
these models predict cosmic density perturbations to be proportional
to the logarithm of the scale. On scales up to the current Hubble
radius $H_{\rm now}^{-1}$, this agrees well with observations of near
scale invariance. However, on much larger length scales of order
$H_{\rm now}^{-1} \exp(10^5)$, perturbations are predicted to be on
the order of one. Of course, this means that the perturbational
treatment breaks down; but it is an indication that black holes may be
created, and thus warrants further investigation.

An attempt to interpret this behaviour was made by
Linde~\cite{Lin86a,Lin86b}. He noted that in the early stages of
inflation, when the strong density perturbations originate, the
quantum fluctuations of the inflaton field are much larger than its
classical decrease per Hubble time. He concluded that therefore there
would always be regions of the inflationary universe where the field
would grow, and so inflation would never end globally (``eternal
inflation''). However, this approach only allows for fluctuations of
the field. One should also consider fluctuations which change the
topology of space--time.  This topology change corresponds to the
formation of a pair of black holes. The pair creation rate can be
calculated using instanton methods, which are well suited to this
non-perturbative problem.

One usually thinks of black holes forming through gravitational
collapse, and so the inflationary era may seem an unlikely place to
look for black holes, since matter will be hurled apart by the rapid
cosmological expansion. However, there are good reasons to expect
black holes to form through the quantum process of pair creation. We
have already pointed out the presence of large quantum fluctuations
during inflation. They lead to strong density perturbations and thus
potentially to spontaneous black hole formation. But secondly, and
more fundamentally, it is clear that in order to pair create {\em any}
object, there must be a force present which pulls the pair apart. In
the case of a virtual electron--positron pair, for example, the
particles can only become real if they are pulled apart by an external
electric field. Otherwise they would just fall back together and
annihilate.  The same holds for black holes: examples in the
literature include their pair creation on a cosmic
string~\cite{HawRos95a}, where they are pulled apart by the string
tension; or the pair creation of magnetically charged black holes on
the background of Melvin's universe~\cite{Gib86},
where the magnetic field prevents them from recollapsing. In our case,
the black holes will be separated by the rapid cosmological expansion
due to the effective cosmological constant. So we see that this
expansion, which we na{\"{\i}}vely expected to prevent black holes
from forming, actually provides just the background needed for their
quantum pair creation.

Since inflation has ended, during the radiation and matter dominated
eras until the present time, the effective cosmological constant was
nearly zero. Thus the only time when black hole pair creation was
possible in our universe was during the inflationary era, when
$\Lambda_{\rm eff}$ was large. Moreover, these black holes are unique
since they can be so small that quantum effects on their evolution are
important. Such tiny black holes could not form from the gravitational
collapse of normal baryonic matter, because degeneracy pressure will
support white dwarfs or neutron stars below the Chandrasekhar limiting
mass.

In the standard semi--classical treatment of pair creation, one finds
two instantons: one for the background, and one for the objects to be
created on the background. From the instanton actions $I_{\rm bg}$
and $I_{\rm obj}$ one calculates the pair creation rate $\Gamma$:
\beq
\Gamma =
\exp \left[ - \left( I_{\rm obj} - I_{\rm bg} \right) \right],
\label{eq-pcr-usual}
\eeq
where we neglect a prefactor. This prescription has been very
successfully used by a number of authors recently
\cite{GarGid94,DowGau94a,DowGau94b,HawHor95,ManRos95,HawRos95b,CalCha96}
for the pair creation of
black holes on various backgrounds.

In this paper, however, we will obtain the pair creation rate through
a somewhat more fundamental, but equivalent procedure: since we have a
cosmological background, we can use the Hartle--Hawking no boundary
proposal~\cite{HarHaw83} for the wave function of the universe.  We
will describe the creation of an inflationary universe by a de~Sitter
type gravitational instanton, which has the topology of a
four--sphere, $S^4$. In this picture, the universe starts out with the
spatial size of one Hubble volume. After one Hubble time, its spatial
volume will have increased by a factor of $e^3 \approx 20$. However,
by the de~Sitter no hair theorem, we can regard each of these $20$
Hubble volumes as having been nucleated independently through
gravitational instantons. With this interpretation, we are allowing
for black hole pair creation, since some of the new Hubble volumes
might have been created through a different type of instanton that has
the topology $S^2 \times S^2$ and thus represents a pair of black
holes in de~Sitter space~\cite{GinPer83}.  Using the framework of the
no boundary proposal (reviewed in Sec.~\ref{sec-nbp}), one can assign
probability measures to both instanton types. One can then estimate
the fraction of inflationary Hubble volumes containing a pair of black
holes by the fraction $\Gamma$ of the two probability measures. This
is equivalent to saying that $\Gamma$ is the pair creation rate of
black holes on a de~Sitter background. We will thus reproduce
Eq.~\rf{eq-pcr-usual}.

In Sec.~\ref{ssec-fixed-cc} we follow this procedure using a
simplified model of inflation, with a fixed cosmological constant,
before going to a more realistic model in
Sec.~\ref{ssec-effective-cc}.  In Sec.~\ref{ssec-discussion} we show
that the usual description of pair creation arises naturally from the
no boundary proposal, and Eq.~\rf{eq-pcr-usual} is recovered.
We find that Planck size black holes can be created in abundance in
the early stages of inflation. Larger black holes, which would form
near the end of inflation, are exponentially suppressed. The
tunnelling proposal~\cite{Vil86}, on the other hand, predicts a
catastrophic instability of de~Sitter space and is unable to reproduce
Eq.~\rf{eq-pcr-usual}.

We then investigate the evolution of black holes in an inflationary
universe. In Sec.~\ref{sec-classical} their classical growth is shown
to correspond to energy-momentum flux across the black hole horizon.
Taking quantum effects into account, we find in Sec.~\ref{sec-quantum}
that the number of neutral black holes that survive into the radiation
era is exponentially small. On the other hand, black holes with a
magnetic charge can also be pair created during inflation. They cannot
decay, because magnetic charge is topologically conserved. Thus they
survive and should still be around today. In Sec.~\ref{sec-charged},
however, we show that such black holes would be too rare to be found
in the observable universe. We summarise our results in
Sec.~\ref{sec-summary}. We use units in which $m_{\rm P} = \hbar = c =
k = 1 $.

\sect{No Boundary Proposal} \label{sec-nbp}

We shall give a brief review; more comprehensive treatments can be
found elsewhere~\cite{EQG}.
According to the no boundary proposal, the quantum state of the
universe is defined by path integrals over Euclidean metrics $ g_{\mu
  \nu} $ on compact manifolds $ M $. From this it follows that the
probability of finding a three--metric $ h_{ij} $ on a spacelike
surface $ \Sigma $ is given by a path integral over all $ g_{\mu \nu}
$ on $ M $ that agree with $ h_{ij} $ on $ \Sigma $. If the spacetime
is simply connected (which we shall assume), the surface $ \Sigma $
will divide $ M $ into two parts, $ M_+ $ and $ M_- $. One can then
factorise the probability of finding $ h_{ij} $ into a product of two
wave functions, $ \Psi_+ $ and $ \Psi_- $.  $ \Psi_+ $ ($\Psi_-$) is
given by a path integral over all metrics $ g_{\mu \nu} $ on the
half--manifold $ M_+ $ ($M_-$) which agree with $ h_{ij} $ on the
boundary $ \Sigma $. In most situations $ \Psi_+ $ equals $ \Psi_- $.
We shall therefore drop the suffixes and refer to $ \Psi $ as the wave
function of the universe. Under inclusion of matter fields, one
arrives at the following prescription:
\beq
\Psi[h_{ij}, \Phi_{\Sigma}] = 
\int \! D(g_{\mu\nu}, \Phi) \,
 \exp \left[ -I(g_{\mu\nu}, \Phi) \right],
 \label{eq-nbp}
\eeq
where $(h_{ij}, \Phi_{\Sigma})$ are the 3-metric and matter field
on a spacelike boundary $\Sigma$ and the path integral is taken over
all compact Euclidean four geometries $g_{\mu\nu}$
that have $\Sigma$ as their
only boundary and matter field configurations
$\Phi$ that are regular on them;
$I(g_{\mu\nu}, \Phi)$ is their action.
The gravitational part of the action is given by
\beq
I_E = -\frac{1}{16\pi} \int_{M_+} \!\!\! d^4\!x\, g^{1/2}(R-2\Lambda) 
      -\frac{1}{8\pi}  \int_{\Sigma} \! d^3\!x\, h^{1/2} K,
\label{eq-action}
\eeq
where $R$ is the Ricci-scalar, $\Lambda$ is the cosmological constant,
and $K$ is the trace of $K_{ij}$, the second fundamental form of the
boundary $\Sigma$ in the metric $g$.

The wave function $ \Psi $ depends on the three--metric $ h_{ij} $ and
on the matter fields $ \Phi $ on $ \Sigma $. It does not, however,
depend on time explicitly, because there is no invariant meaning to
time in cosmology. Its independence of time is expressed by the fact
that it obeys the Wheeler--DeWitt equation. We shall not try to solve
the Wheeler--DeWitt equation directly, but we shall estimate $ \Psi $
from a saddle point approximation to the path integral.

We give here only a brief summary of this semi--classical method; the
procedure will become clear when we follow it through in the following
section. We are interested in two types of inflationary universes: one
with a pair of black holes, and one without. They are characterised
by spacelike sections of different topology. For each of
these two universes, we have to find a classical Euclidean solution to
the Einstein equations (an instanton), which can be analytically
continued to match a boundary $ \Sigma $ of the appropriate
topology.  We then calculate the Euclidean actions $I$ of the two
types of solutions.  Semiclassically, it follows from Eq.~\rf{eq-nbp}
that the wave function is given by
\beq
\Psi = \exp \left( -I \right),
\label{eq-nbp-sca}
\eeq
where we neglect a prefactor.
We can thus assign a probability measure to each type of universe:
\beq
P = \left| \Psi \right|^2 = \exp \left( -2I^\re \right),
\label{eq-prob-measure}
\eeq
where the superscript `Re' denotes the real part.
As explained in the introduction, the ratio of the two probability
measures gives the rate of black hole pair creation on an
inflationary background, $\Gamma$.

\sect{Creation of Neutral Black Holes} \label{sec-creation}

The solutions presented in this section are discussed much more
rigorously in an earlier paper~\cite{BouHaw95}.  We shall assume
spherical symmetry.  Before we introduce a more realistic inflationary
model, it is helpful to consider a simpler situation with a fixed
positive cosmological constant $ \Lambda $ but no matter fields. We
can then generalise quite easily to the case where an effective
cosmological ``constant'' arises from a scalar field.

\subsection{Fixed Cosmological Constant} \label{ssec-fixed-cc}

\subsubsection{The de~Sitter solution}
 
First we consider the case without black holes: a homogeneous
isotropic universe.  Since $\Lambda > 0$, its
spacelike sections will simply be round three--spheres. The wave
function is given by a path integral over all metrics on a
four--manifold $ M_+ $ bounded by a round three--sphere $ \Sigma $ of
radius $ a_{\Sigma} $.  The corresponding saddle point solution is
the de~Sitter space--time.  Its Euclidean metric is that of a round
four--sphere of radius $\sqrt{3/\Lambda}$:
\beq
ds^2 = d\tau^2 + a(\tau)^2 d\Omega_3^2,
\eeq
where $\tau$ is Euclidean time, $d\Omega_3^2$ is the metric on the
round three--sphere of unit radius, and
\beq
a(\tau) = \sqrt{\frac{3}{\Lambda}} \sin \sqrt{\frac{\Lambda}{3}} \tau.
\label{eq-a-LS4}
\eeq

For $ a_{\Sigma}=0 $, the saddle point metric will only be a single
point. For $0 < a_{\Sigma} < \sqrt{3/\Lambda} $ it will be part of the
Euclidean four--sphere , and when $ a_{\Sigma} = \sqrt{3 / \Lambda} $,
the saddle point metric will be half the four--sphere.  When $
a_{\Sigma} > \sqrt{3/\Lambda} $ there will be no real Euclidean metric
which is a solution of the field equations with the given boundary
conditions. However, we can regard Eq.~\rf{eq-a-LS4} as a function on
the complex $\tau$--plane. On a line parallel to the imaginary
$\tau$--axis defined by $\tau^\re = \sqrt{\frac{3}{\Lambda}} \,
\frac{\pi}{2} $, we have
\beq 
\left. a(\tau) \right|_{\tau^{\rm Re}=
                        \sqrt{\frac{3}{\Lambda}} \frac{\pi}{2}} = 
 \sqrt{\frac{3}{\Lambda}} \cosh \sqrt{\frac{\Lambda}{3}} \tau^{\rm Im}.
\eeq
This describes a Lorentzian de~Sitter hyperboloid, with $\tau^\im$
serving as a Lorentzian time variable.  One can thus construct a
complex solution, which is the analytical continuation of the
Euclidean four--sphere metric.  It is obtained by choosing a contour
in the complex $\tau$--plane from $0$ to $\tau^\re =
\sqrt{\frac{3}{\Lambda}} \, \frac{\pi}{2} $ and then parallel to the
imaginary $\tau$--axis. One can regard this complex solution as half
the Euclidean four--sphere joined to half of the Lorentzian de~Sitter
hyperboloid
  \begin{figure}[htb] 
   \hspace{.1\textwidth}
   \vbox{
    \epsfxsize=.8\textwidth
    \epsfbox{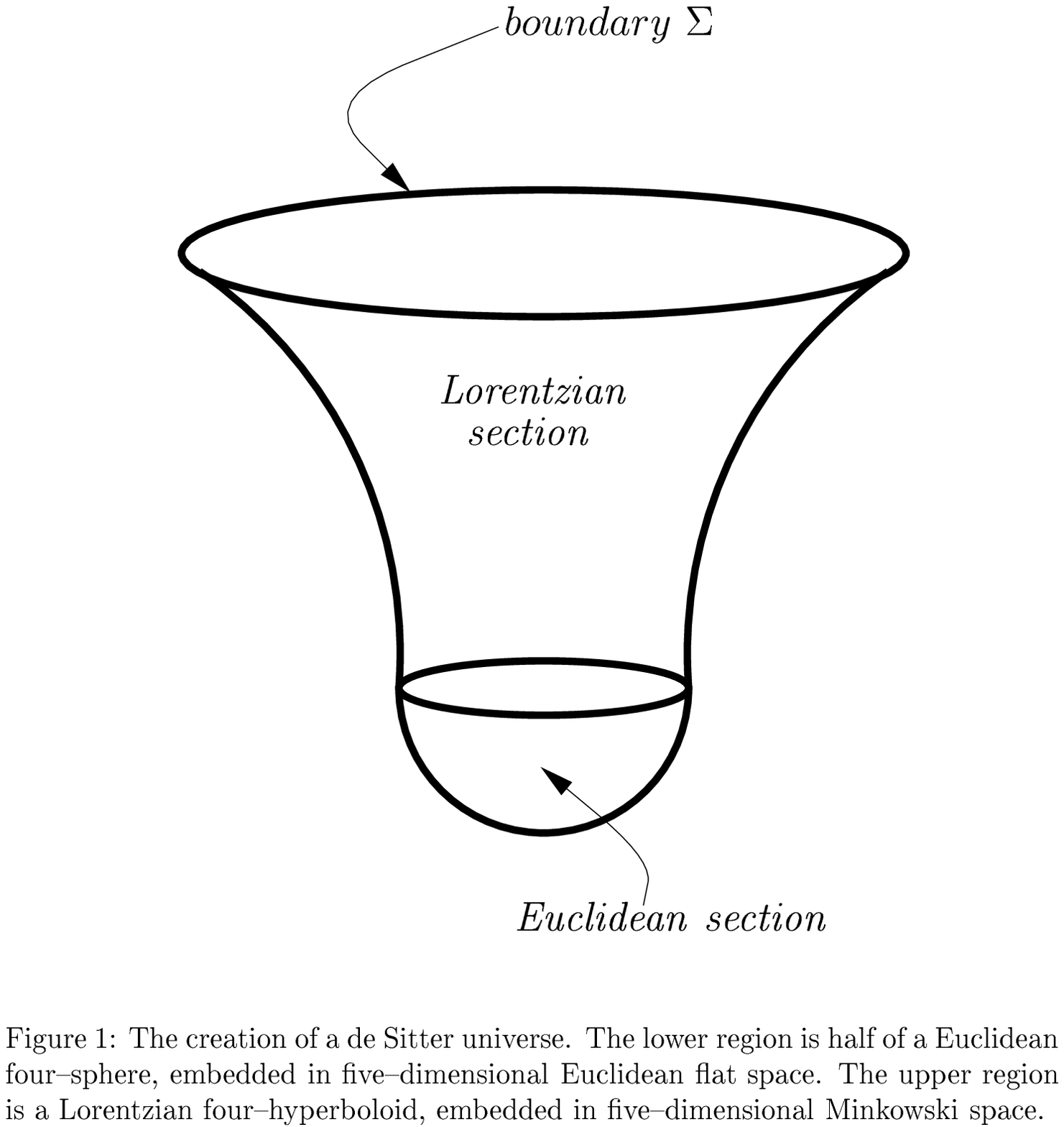}
    }
   \label{fig-creation}
  \end{figure}
(Fig.~1). 

The Lorentzian
part of the metric will contribute a purely imaginary term
to the action. This will affect the phase of the wave function but not
its amplitude. The real
part of the action of this complex saddle point metric will be the
action of the half Euclidean four--sphere:
\beq
I^\re_{\rm de\ Sitter} = - \frac{3\pi}{2\Lambda}.
\eeq
Thus the magnitude of the
wave function will still be $ e^{3 \pi / 2 \Lambda} $, corresponding
to the probability measure
\beq
P_{\rm de\ Sitter} = \exp \left( \frac{3\pi}{\Lambda} \right).
\label{eq-prob-desitter}
\eeq

\subsubsection{The Schwarzschild--de~Sitter solution}

We turn to the case of a universe containing a pair of black holes.
Now the cross sections $ \Sigma $ have topology $ S^2 \times S^1 $.
Generally, the radius of the $ S^2 $ varies along the $ S^1 $.  This
corresponds to the fact that the radius of a black hole immersed in
de~Sitter space can have any value between zero and the radius of the
cosmological horizon.  The minimal two--sphere corresponds to the
black hole horizon, the maximal two--sphere to the cosmological
horizon.  The saddle point solution corresponding to this topology is
the Schwarzschild--de~Sitter universe.  However, the Euclidean section
of this spacetime typically has a conical singularity at one of its
two horizons and thus does not represent a regular instanton. This is
discussed in detail in the Appendix. There we show that the only
regular Euclidean solution is the degenerate case where the black hole
has the maximum possible size. It is also known as the Nariai solution
and given by the topological product of two round two--spheres:
\beq
ds^2 = d\tau^2 + a(\tau)^2 dx^2 + b(\tau)^2 d\Omega_2^2,
\eeq
where $x$ is identified with period $2\pi$, $ d\Omega_2^2 = d\theta^2
+ \sin^2\! \theta\: d\varphi^2 $, and
\beq
a(\tau) = \sqrt{\frac{1}{\Lambda}} \sin \sqrt{\Lambda} \tau, \:
b(\tau) = \sqrt{\frac{1}{\Lambda}} = {\it const}.
\eeq

In this case the radius $b$ of the $ S^2 $ is constant in the $ S^1 $
direction. The black hole and the cosmological horizon have equal
radius $1/\sqrt{\Lambda}$ and no conical singularities are present.
Thus, by requiring the smoothness of the Euclidean solution, the
instanton approach not only tells us about probability measures, but
also about the size of the black hole. There will be no saddle point
solution unless we specify $ b_{\Sigma}=1/\sqrt{\Lambda} $.  We are
then only free to choose the radius $a_{\Sigma}$ of the one--sphere on
$\Sigma$. For this variable,
the situation is similar to the de~Sitter case.
There will be real Euclidean saddle point metrics on $ M_+ $ for $
a_{\Sigma} \leq 1 / \sqrt{\Lambda} $. For larger $a_{\Sigma}$ there
will again be no Euclidean saddle point, but we find that
\beq
\left. a(\tau) \right|_{\tau^{\rm Re}=
                        \sqrt{\frac{1}{\Lambda}} \frac{\pi}{2}} = 
 \sqrt{\frac{1}{\Lambda}} \cosh \sqrt{\Lambda} \tau^{\rm Im}.
\eeq
This corresponds to the Lorentzian section of the degenerate
Schwarzschild--de~Sitter spacetime, in which the $S^1$
expands rapidly,
while the two--sphere (and therefore the black hole radius) remains
constant. Again we can
construct a complex saddle point, which can be regarded as
half a Euclidean $ S^2 \times S^2 $ joined to half of the Lorentzian
solution. The real part of the action will be the action of the half
of a Euclidean $ S^2 \times S^2 $:
\beq
I^\re_{\rm SdS} = -\frac{\pi}{\Lambda}.
\eeq
The corresponding probability measure is
\beq
P_{\rm SdS} = \exp \left( \frac{2\pi}{\Lambda} \right).
\label{eq-prob-nariai} 
\eeq
We divide this by the probability measure~\rf{eq-prob-desitter} for a
universe without black holes to obtain the pair creation rate of black
holes in de~Sitter space:
\beq
\Gamma =
\frac{P_{\rm SdS}}{P_{\rm de\ Sitter}} =
       \exp \left( - \frac{\pi}{\Lambda} \right).
\eeq
Thus the probability for pair creation is very low,
unless $ \Lambda $ is close to the Planck value, $ \Lambda = 1 $.

\subsection{Effective Cosmological Constant}
  \label{ssec-effective-cc}

Of course the real universe does not have a cosmological constant of
order the Planck value. However, in inflationary cosmology it is
assumed that the universe starts out with a very large effective
cosmological constant, which arises from the potential $ V $ of a
scalar field $ \phi $. The exact form of the potential is not
critical. So for simplicity we chose $ V $ to be the potential of a
field with mass $ m $, but the results would be similar for a $
\lambda \phi^4 $ potential. To account for the observed fluctuations
in the microwave background~\cite{COBE}, $ m $ has to be on the order
of $ 10^{-5} $ to $ 10^{-6} $~\cite{HalHaw85}.  The wave function $
\Psi $ will now depend on the three--metric $ h_{ij} $ and the value
of $ \phi $ on $ \Sigma $. By a gauge choice one can take $ \phi $ to
be constant on $ \Sigma $, and we shall do so for simplicity. For $
\phi > 1 $ the potential acts like an effective cosmological constant
\beq
\Lambda_{\rm eff}(\phi) = 8 \pi V(\phi).
\eeq

One proceeds in complete analogy to the fixed cosmological constant
case.  For small three--geometries and $ \phi > 1 $, there will be
real Euclidean metrics on $ M_+ $, with $ \phi $ almost constant.  If
the three--geometries are rather larger, there will again not be any
real Euclidean saddle point metrics. There will however be complex
saddle points. These can again be regarded as a Euclidean solution
joined to a Lorentzian solution, although neither the Euclidean nor
Lorentzian metrics will be exactly real.  Apart from this subtlety,
which is dealt with in Ref.~\cite{BouHaw95}, the saddle point
solutions are similar to those for a fixed cosmological constant, with
the time--dependent $\Lambda_{\rm eff}$ replacing $ \Lambda $.  The
radius of the pair created black holes will now be given by
$1/\sqrt{\Lambda_{\rm eff}}$.  As before, the magnitude of the wave
function comes from the real part of the action, which is determined
by the Euclidean part of the metric. This real part will be
\beq
I^\re_{S^3} = - \frac{3 \pi}{2 \Lambda_{\rm eff}(\phi_0)}
\eeq
in the case without black holes, and
\beq
I^\re_{S^2 \times S^1} = - \frac{\pi}{\Lambda_{\rm eff}(\phi_0)}
\eeq
in the case with a black hole pair. Here $ \phi_0 $ is the value of $
\phi $ in the initial Euclidean region. Thus the pair creation rate is
given by
\beq
\Gamma = 
\frac{P_{S^2 \times S^1}}{P_{S^3}} =
         \exp \left[ - \frac{\pi}{\Lambda_{\rm eff}(\phi_0)} \right].
\label{eq-suppression}
\eeq

\subsection{Discussion} \label{ssec-discussion}

Let us interpret this result.  Since $ 0 < \Lambda_{\rm eff} \lesssim
1 $, we get $\Gamma < 1$ and so black hole pair creation is
suppressed. In the early stages of inflation, when $\Lambda_{\rm eff}
\approx 1$, the suppression is week, and black holes will be
plentifully produced. However, those black holes will be very small,
with a mass on the order of the Planck mass.  Larger black holes,
corresponding to lower values of $\Lambda_{\rm eff}$ at later stages
of inflation, are exponentially suppressed.  We shall see in the
following two sections that the small black holes typically evaporate
immediately, while sufficiently large ones grow with the horizon and
survive long after inflation ends (that is, long in early universe
terms).

We now understand how the standard prescription for pair creation,
Eq.~\rf{eq-pcr-usual}, arises from this proposal: by
Eq.~\rf{eq-prob-measure},
\beq
\Gamma = 
\frac{P_{S^2 \times S^1}}{P_{S^3}} =
\exp \left[ - \left( 2I^\re_{S^2 \times S^1}-2I^\re_{S^3}
  \right) \right],
\eeq
where $I^\re$ denotes the real part of the Euclidean action of a
complex saddle point solution. But we have seen that this is equal to
half of the action of the complete Euclidean solution. Thus $I_{\rm
  obj} = 2I^\re_{S^2 \times S^1}$ and $I_{\rm bg} = 2I^\re_{S^3}$, and
we recover Eq.~\rf{eq-pcr-usual}.

The prescription for the wave function of the
universe has long been one of the central, and arguably one of the
most disputed issues in quantum cosmology.  According to Vilenkin's
tunnelling proposal~\cite{Vil86}, $\Psi$ is given by $e^{+I}$ rather
than $e^{-I}$.  This choice of sign is inconsistent with
Eq.~\rf{eq-pcr-usual}, as it leads to the inverse result for the pair
creation rate: $\Gamma_{\rm TP} = 1/\Gamma_{\rm NBP}$.  In our case,
we would get $\Gamma_{\rm TP} = \exp ( + \pi/\Lambda_{\rm eff})$.
Thus black hole pair creation would be enhanced, rather than
suppressed. De~Sitter space would be catastrophically unstable to the
formation of black holes.  Since the radius of the black holes is
given by $1/\sqrt{\Lambda_{\rm eff}}$, the black holes would be more
likely the larger they were. Clearly, the tunnelling proposal cannot
be maintained.  On the other hand, Eq.~\rf{eq-suppression}, which was
obtained from the no boundary proposal, is physically very reasonable.
It allows topological fluctuations near the Planckian regime, but
suppresses the formation of large black holes at low energies.  Thus
the consideration of the cosmological pair production of black holes
lends strong support to the no boundary proposal.

\sect{Classical Evolution} \label{sec-classical}

We shall now consider neutral black holes created at any value
$\phi_0>1$ of the scalar field and analyse the different effects on
their evolution. 
Before we take quantum effects into account, we shall display the
classical solution for a universe containing a pair of black holes.
We shall demonstrate explicitly that it behaves according to the First
Law of black hole mechanics.

With a rescaled inflaton potential
\beq
V(\phi) = \frac{1}{8\pi} m^2 \phi^2,
\eeq
the effective cosmological constant will be
\beq
\Lambda_{\rm eff} = m^2 \phi^2.
 \label{eq-eff-cc}
\eeq
In the previous section we learned that the black hole radius remains
constant, at $1/\sqrt{\Lambda}$, in the Lorentzian regime. But this
was for the simple model with fixed $\Lambda$. The effective
cosmological constant in Eq.~\rf{eq-eff-cc} is slightly time
dependent. Thus we might expect the black hole size to change during
inflation.

Indeed, for $ \frac{\pi}{2m \phi_0} < t <
\frac{\phi_0}{m} $, approximate Lorentzian solutions are given
by~\cite{BouHaw95} 
\beqa
\phi (t) & = & \phi_0 - mt,
\label{eq-phi}
\\
a (t)    & = & \frac{1}{m \phi_0} 
     \cosh \left[ m \! \int_0^t \!\! \phi(t') \, dt' \right],
\label{eq-a}
\\
b (t)    & = & \frac{1}{m \phi(t)},
\label{eq-b}
\\
ds^2     & = & -dt^2 + a(t)^2 dx^2 + b(t)^2 d\Omega_2^2.
\label{eq-KS}
\eeqa
Since we are dealing with a degenerate solution,
the radii $ r_{\rm b} $
and $ r_{\rm c} $ of the black
hole and cosmological horizons are equal:
\beq
r_{\rm b} = r_{\rm c} = b(t).
\eeq
According Eq.~\rf{eq-b} they will expand slowly together during
inflation as the scalar field rolls down to the minimum of the
potential $ V $ and the effective cosmological constant decreases.
At the end of inflation they will be approximately equal to
$ m^{-1} $. 

One can think of this increase of the horizons as a classical effect,
caused by a flow of energy--momentum across them. If the scalar field
were constant, its energy--momentum tensor would act exactly like a
cosmological constant. The flow of energy--momentum across the horizon
would be zero. However, the scalar field is not constant but is
rolling down hill in the potential to the minimum at $ \phi = 0
$. This means that there is an energy--momentum flow across the
horizon equal to
\beq
\dot{M} = A\; T_{ab} l^a l^b = \frac{1}{\phi^2}, \label{eq-cl-flow}
\eeq
where $ A = 4 \pi b^2 $ is the horizon area,
$ T_{ab} $ is the energy--momentum tensor for the massive scalar
field, given by
\beq
T_{ab} =   \frac{1}{4\pi} \partial_a \phi \, \partial_b \phi
         - \frac{1}{8\pi} g_{ab} \left( \partial_c \phi \, \partial^c
            \! \phi + m^2 \phi^2 \right),
\eeq
and $ l^a $ is a null vector tangent to the horizon:
\beq
l^a =   \frac{\partial}{\partial t}
      + \frac{1}{a} \frac{\partial}{\partial x}.
\eeq

One would expect the horizons to respond to this flow of energy across
them by an increase in area according to the First Law of black hole
mechanics~\cite{BarCar73}:
\beq
\dot{M} = \frac{\kappa}{8\pi} \dot{A}, \label{eq-first-law}
\eeq
where $ \kappa $ is the surface gravity of the horizon.
We will show that this equation is indeed satisfied if the horizon
growth is given by Eq.~\rf{eq-b}.

The values of $ \kappa $ for general Schwarzschild--de~Sitter
solutions are derived in the Appendix.
Due to the slow change of the effective cosmological constant we can
approximate the surface gravity at any time $ t $ in our model by the
surface gravity in the model with a fixed cosmological constant $
\Lambda = \Lambda_{\rm eff}(t) $. In the degenerate case which we are
considering now, $ \kappa $ will thus be given by
\beq
\kappa = \sqrt{\Lambda_{\rm eff}} = m\phi.
\eeq
Eq.~\rf{eq-first-law} becomes
\beq
\dot{M} = \frac{m\phi}{8\pi} \;
          \frac{d}{dt} \! \left( \frac{4\pi}{m^2 \phi^2} \right)
        = \frac{1}{\phi^2},
\eeq
which agrees with Eq.~\rf{eq-cl-flow}.

It should be pointed out that this calculation holds not only for the
black hole horizon, but also for the cosmological horizon. Moreover,
an analogous calculation is possible for the cosmological horizon in
an ordinary inflationary universe without black holes. Thus, in
hindsight we understand the slow growth of the cosmological horizon
during inflation as a manifestation of the First Law of black hole
mechanics.

\sect{Quantum Evolution} \label{sec-quantum}

So far we have been neglecting the quantum properties of the
inflationary spacetime presented above. It is well known that in a
Schwarzschild--de~Sitter universe, radiation is emitted both by the
black hole and the cosmological horizon~\cite{GibHaw77b}. To treat this
properly, one should include the one--loop effective action of all the
low mass fields in the metric $ g_{\mu \nu} $. By using a
supersymmetric theory one might avoid divergences in the one--loop
term, but it would still be impossibly difficult to calculate in any
but very simple metrics. Instead, we shall use an approximation in
which the black hole and cosmological horizons radiate thermally with
temperatures
\beq
T_{\rm b} = \frac{\kappa_{\rm b}}{2\pi},\;\;
T_{\rm c} = \frac{\kappa_{\rm c}}{2\pi}.
\label{eq-temperature}
\eeq

This quantum effect must also be included in the calculation of the
energy flow across the horizons. For the saddle point metric,
Eq.~\rf{eq-KS}, it has no consequence: in the Nariai solution the
black hole and cosmological horizons have the same radius and surface
gravity. Thus they radiate at the same rate. That means they will be
in thermal equilibrium. The black holes will not evaporate, because
they will be absorbing as much as they radiate. Instead, their
evolution will be governed by the classical growth described above.

However, the Nariai metric is an idealization. (Strictly
speaking, it does not even contain a black hole, but rather two
acceleration horizons.) Due to quantum fluctuations there will be
small deviations from the saddle point solution, corresponding to a
Schwarzschild--de~Sitter spacetime which is not quite degenerate. The
radius $ b $ of the two--sphere will not be exactly constant along the
$ S^1 $, but will have a maximum $ r_{\rm c} $ and a minimum $ r_{\rm
  b} $, which we identify with the cosmological and the black hole
horizons, respectively. (The other black hole horizon of the pair will
lie beyond the cosmological horizon and will not be visible in our
universe.) Since the black hole horizon is slightly smaller, it will
have a higher temperature than the cosmological horizon. Therefore the
black hole will radiate more than it receives.  There will thus be a
net transfer of energy from the smaller horizon to the larger one.
This will cause the larger horizon to grow faster, and the smaller one
to shrink until the black hole vanishes completely. We show below that
black holes can still grow with the cosmological horizon until the end
of inflation, if they are either created sufficiently large, or start
out very nearly degenerate.  However, we shall see that none of these
conditions is easily satisfied.

Black holes created during the final stages of inflation will survive
until the end of inflation simply because they will be relatively
large and cold. One can estimate the minimum size they must have by
treating them as evaporating Schwarzschild black holes, which have a
lifetime on the order of $M^3$, where $M$ is the mass at which the
black hole is created. In terms of the value of the scalar field at
creation, $ M \approx b_0 =  (m \phi_0)^{-1} $. Inflation ends after a
time of $ \phi_0 / m $.  Therefore black holes created at $ \phi_0
\leq m^{-1/2} $ will certainly survive until the end of inflation.
They would continue to grow slowly during the
radiation era, until the temperature of the radiation falls below that
of the black holes. They will then start to evaporate.
By Eq.~\rf{eq-suppression}, however, such black holes will be
suppressed by a factor of
\beq
\Gamma = \exp \left(  - \pi m^{-1} \right).
\label{eq-suppression-safe}
\eeq

One must therefore investigate the possibility that the small black
holes, which can be created in abundance, start out so nearly
degenerate that they will grow with the cosmological horizon until
they have reached the ``safe'' size of $ m^{-1/2} $. We need to
determine how nearly equal the horizon sizes have to be initially
so that the black hole survives until the end of inflation.
 
If we take the thermal radiation into account, the flow across the
horizons now consists of two parts: the classical term due to the
energy flow of the scalar field, as well as the net radiation energy
transfer, given by Stefan's law. Applying the First Law of black hole
mechanics to each horizon, we get:
\begin{eqnarray}
\frac{\kappa_{\rm b}}{8\pi} \dot{A_{\rm b}}
        & = & m^2 r_{\rm b}^2 - \left(
        \sigma A_{\rm b} T_{\rm b}^4 -
        \sigma A_{\rm c} T_{\rm c}^4 \right),
\label{eq-evol1a}
\\
\frac{\kappa_{\rm c}}{8\pi} \dot{A_{\rm c}}
        & = & m^2 r_{\rm c}^2 + \left(
        \sigma A_{\rm b} T_{\rm b}^4 - 
        \sigma A_{\rm c} T_{\rm c}^4 \right),
\label{eq-evol1b}
\end{eqnarray}
where $ \sigma = \pi^2 / 60 $ is the Stefan--Boltzmann constant.
Using Eq.~\rf{eq-temperature}, we obtain two coupled differential
equations for the horizon radii:
\begin{eqnarray}
\dot{r}_{\rm b} & = & \frac{1}{\kappa_{\rm b} r_{\rm b}} \left[
        m^2 r_{\rm b}^2 - \frac{1}{240 \pi} \left(
        r_{\rm b}^2 \kappa_{\rm b}^4 -
        r_{\rm c}^2 \kappa_{\rm c}^4 \right)
        \right],
\label{eq-evol2a}
\\
\dot{r}_{\rm c} & = & \frac{1}{\kappa_{\rm c} r_{\rm c}} \left[ 
        m^2 r_{\rm c}^2 + \frac{1}{240 \pi} \left(
        r_{\rm b}^2 \kappa_{\rm b}^4 -
        r_{\rm c}^2 \kappa_{\rm c}^4 \right)
        \right].
\label{eq-evol2b}
\end{eqnarray}

The exact functional relation between the surface gravities and the
horizon radii is generally non--trivial. However, the above evolution
equations can be simplified if one takes into account that a nucleated
black hole pair must be very nearly degenerate if it is to survive
until the end of inflation. We can therefore write
\beq
r_{\rm b} (t) = b(t) \left[ 1 - \eps(t) \right], \;\;
r_{\rm c} (t) = b(t) \left[ 1 + \eps(t) \right],
\eeq
with $ \eps_0 \ll 1 $.
The surface gravities can also be expressed in terms of $ \eps $
(see the Appendix); to first order they are given by:
\beq
\kappa_{\rm b} = \frac{1}{b(t)}
 \left[1+\frac{2}{3} \eps(t) \right],\;\;  
\kappa_{\rm c} = \frac{1}{b(t)}
 \left[1-\frac{2}{3} \eps(t) \right].
\eeq  
We assume that $ b(t) $ behaves as in
Eq.~\rf{eq-b} for the Nariai solution as long as $ \eps(t) \ll 1 $.
Eqs.~\rf{eq-evol2a}, \rf{eq-evol2b} are then identically satisfied to
zeroth order in $ \eps $. In the first order, they give an evolution
equation for $ \eps $:
\beq
\dot{\eps} = \left( \frac{2}{3} m^2 b +
             \frac{1}{180\pi} \frac{1}{b^3} \right) \eps.
\eeq
This equation can be integrated to give
\beq
\eps(t) = \eps_0 \left( \frac{\phi}{\phi_0} \right)^{2/3}
          \exp \left[ \frac{1}{720 \pi} m^2 \phi_0^4 \left( 1 -
          \left( \frac{\phi}{\phi_0} \right)^4 \right) \right].
\eeq

For the unsuppressed, Planck size black holes we have $ \phi_0 =
m^{-1} $. If they grow with the horizon, they will reach the safe
size, which corresponds to $ \phi = m^{-1/2} $, after a time
\beq
t_{\rm safe} = m^{-2} \left( 1 - m^{1/2} \right).
\eeq
If a Planck size black hole is to have survived until this time,
i.e. if $ \eps(t_{\rm safe}) \leq 1 $, then the initial difference
in horizon sizes may not have been larger than
\beq
\eps_0^{\rm max} =
         m^{1/3} \exp \left[ - \frac{m^{-2}}{720 \pi}
         \left( 1 - m^2 \right) \right].
\eeq

The probability $ P(\eps_0 \leq \eps_0^{\rm max}) $ that the two
horizons start out so nearly equal obviously depends on the
distribution of the initial sizes of the two horizons. The
semi--classical treatment of the quantum fluctuations which cause the
geometry to differ from the degenerate case, for general values of the
effective cosmological constant, is an interesting problem by itself,
and beyond the scope of this paper. We hope to return to it in a
forthcoming paper on complex solutions in quantum cosmology. However,
here we are working at the Planck scale, so that the semi--classical
approximation will break down anyway. It therefore seems reasonable to
assume that the initial sizes of the horizons are distributed roughly
uniformly between zero and a few Planck lengths.  This means that
\beq
P \left( \eps_0 \leq \eps_0^{\rm max} \right) \approx
   \eps_0^{\rm max} \approx
   \exp \left( - \frac{m^{-2}}{720 \pi} \right).
\eeq
Since $ m \approx 10^{-6} $ we conclude by comparison with the
suppression factor~\rf{eq-suppression-safe} that it is considerably
less efficient to create Planck size black holes which would grow to
the safe size than just to create the large black holes. Both
processes, however, are exponentially suppressed.

\sect{Charged Black Holes} \label{sec-charged}

Although nearly all neutral black holes will evaporate during
inflation, those that have a magnetic charge won't be able to because
there are no magnetically charged particles for them to radiate.
They can only evaporate down to the minimum mass necessary to support
their magnetic charge.
Let us therefore introduce a Maxwell term in the action and re-examine
the pair creation of primordial black holes:
\beq
I   = -\frac{1}{16\pi} \int_{M_+} \! d^4\!x\, g^{1/2}
      \left( R - 2\Lambda - F_{\mu\nu} F^{\mu\nu} \right) 
      -\frac{1}{8\pi}  \int_{\Sigma} d^3\!x\, h^{1/2} K.
\eeq
The $ S^2 \times S^2 $ bubbles in spacetime foam can now carry magnetic
flux. The action of the Maxwell field will reduce their probability
with respect to neutral bubbles. Thus magnetically charged black holes
will also be pair created during inflation, though in smaller numbers
than neutral black holes. However, once created, they can disappear
only if they meet a black hole with the opposite charge, which is
unlikely. So they should still be around today.

We shall estimate the number of primordial charged black holes present
in the observable universe. For the purpose of calculating the pair
creation rate during inflation, we can use the solutions for
a fixed cosmological constant, since $ \Lambda_{\rm eff} $ changes
slowly.  There exists a three--parameter family of Lorentzian charged
Schwarzschild--de~Sitter solutions.  They are usually called
Reissner--Nordstr\"om--de~Sitter solutions and labeled by the charge $
q $ and the ``mass'' $ \mu $ of the black hole, and by the
cosmological constant $\Lambda$:
\begin{equation}
ds^2 = - U(r) dt^2 + U(r)^{-1} dr^2 + r^2 d\Omega_2^2,
  \label{eq-RNdS}
\end{equation}
where
\beq
U(r) = 1 - \frac{2 \mu}{r} + \frac{q^2}{r^2} - \frac{1}{3} \Lambda r^2.
\eeq
We are interested in the cases where the black holes are magnetically,
rather than electrically charged. Then the Maxwell field is given by
\beq
F = q \sin \theta \, d\theta \wedge d\phi.
\eeq

In an appropriate region of the parameter space, $ U $ has three
positive roots, which we denote, in ascending order, by $ r_{\rm i} $,
$ r_{\rm o} $ and $ r_{\rm c} $ and interpret as the inner and outer
black hole horizons and the cosmological horizon. They can serve as an
alternative parametrisation of the solutions. For general values of
$q$, $\mu$ and $\Lambda$ the metric~\rf{eq-RNdS} has no regular
Euclidean section. The black holes which can be pair created through
regular instantons lie on three intersecting hypersurfaces in the
parameter space~\cite{MelMos89}, as seen in
  \begin{figure}[htb]
   \hspace{.1\textwidth}
   \vbox{
    \epsfxsize=.8\textwidth
    \epsfbox{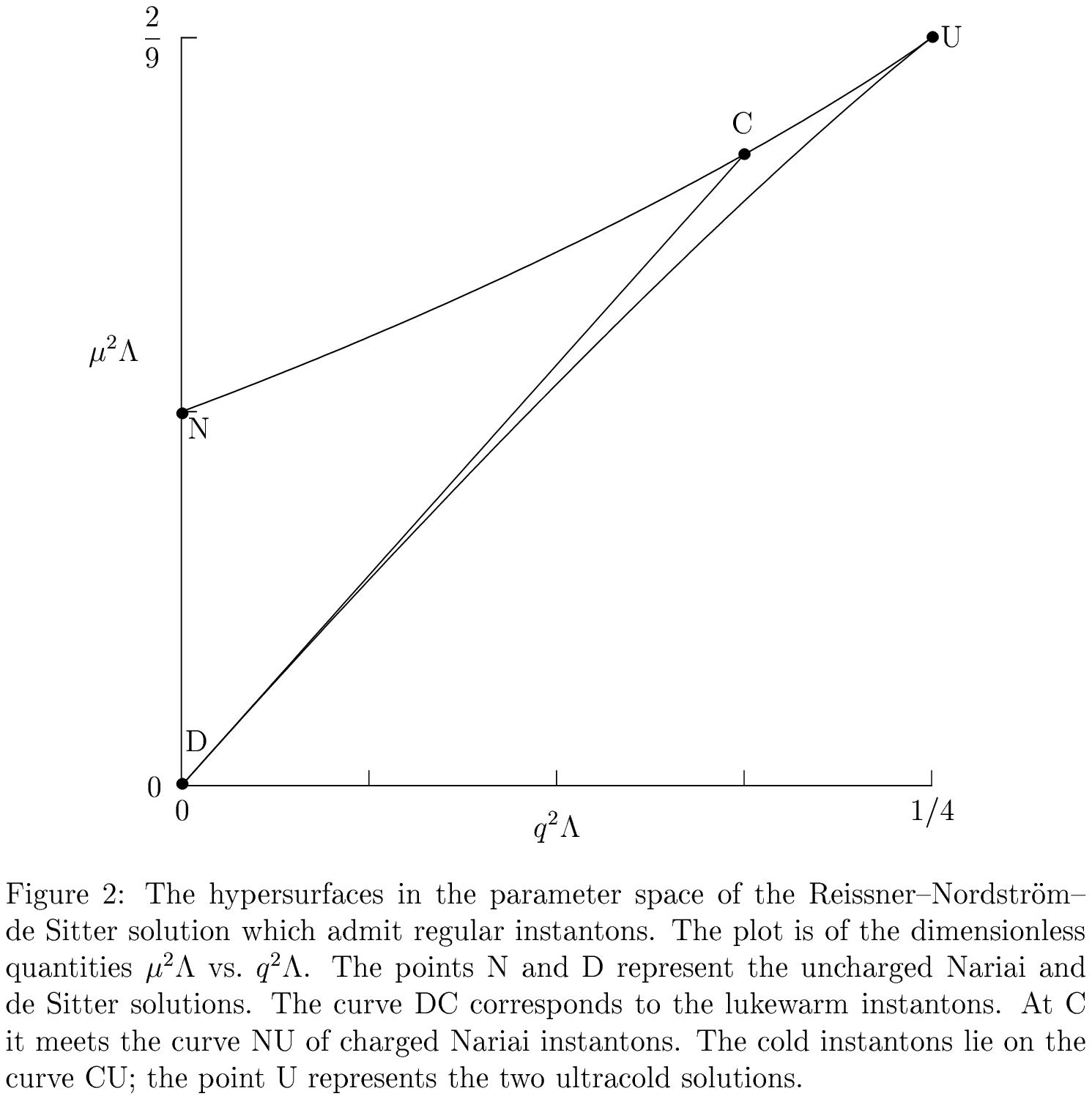} 
    }
   \label{fig-instantons}
  \end{figure}
Fig.~2. They are
called the {\em cold}, the {\em lukewarm} and the {\em charged Nariai}
solutions. In the cold case, the instanton is made regular by setting
$r_{\rm i} = r_{\rm o}$, which corresponds to an extremal black hole.
The lukewarm hypersurface is characterised by the condition $q=\mu$.
It corresponds to a non--extremal black hole which has the same
surface gravity and temperature as the cosmological horizon. This
property is shared by the charged Nariai solution, which has $r_{\rm
  o} = r_{\rm c}$. For $q=0$, its mass is given by $\mu =
1/(3\sqrt{\Lambda})$ and it coincides with the neutral Nariai universe
we discussed earlier. For larger charge and mass, it is still the
direct product of two round two--spheres, however with different
radii. The cold, lukewarm and de~Sitter solutions all coincide for
$q=\mu=0$.  The largest possible mass and charge for the lukewarm
solution is $ q = \mu = 3/(4\sqrt{3\Lambda}) $, where it coincides
with the charged Nariai solution. The largest possible mass and charge
for any regular instanton is attained at the point where the charged
Nariai and the cold hypersurfaces meet. This {\em ultracold} case has
$ q = 1/(2\sqrt{\Lambda}) $ and $ \mu = 2/(3\sqrt{2\Lambda}) $. It
admits two distinct solutions of different action. All of these
solutions are presented and discussed in detail in the comprehensive
paper by Mann and Ross~\cite{ManRos95}, where the actions are
calculated as well. We shall now apply these results in the context of
inflation.

Let us consider an inflationary scenario in which the effective
cosmological constant $\Lambda_{\rm eff} = m \phi$ starts out near the
Planck value and then decreases slowly.
As in the neutral case, one would expect the creation of charged black
holes to be least suppressed for large $\Lambda$, i.e.\ at the
earliest stage of inflation. However, unlike the neutral case a
magnetically charged black hole cannot be arbitrarily small since it
must carry at least one unit of magnetic charge:
\beq
q_0 = \frac{1}{2e_0},
\eeq
where $ e_0 = \sqrt{\alpha} $
is the unit of electric charge, and $ \alpha \approx 1/137 $
is the fine structure constant.
In the following we shall only consider black holes with $q=q_0$,
since they are they first to be created, and
since more highly charged black holes are exponentially suppressed
relative to them. We see from 
  \begin{figure}[htb]
   \hspace{.1\textwidth}
   \vbox{
    \epsfxsize=.8\textwidth
    \epsfbox{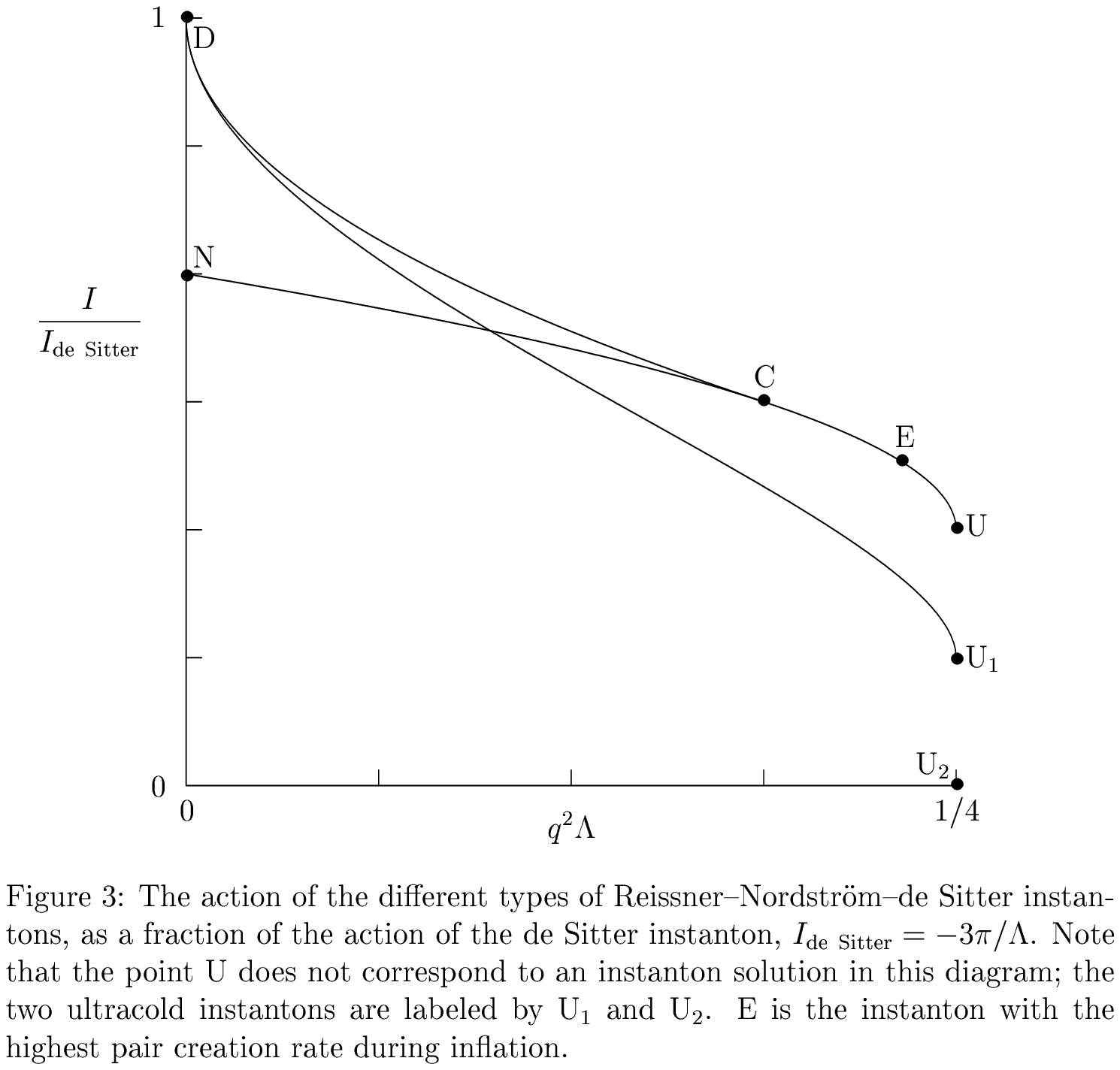} 
    }
   \label{fig-actions}
  \end{figure}
Fig.~3 that pair creation first becomes possible
through the ultracold instanton U$_1$,
when $ \Lambda_{\rm eff} $ has decreased to the value of
$ \Lambda^{\rm U} = 1/(4 q_0^2) = \alpha $.
Since this solution has relatively high action,
however, black hole production becomes more efficient at a
slightly later time. It will then occur mainly through
the charged Nariai instanton, which has lower action than the cold
black hole. When
\beq
\Lambda_{\rm eff} = \Lambda^{\rm E}
                  = \left( 4 \sqrt{3} - 6 \right) \alpha,
\eeq
the pair creation rate reaches its peak and is given by
\beq
\Gamma =
\exp \left[ - \left( I^{\rm E} - I_{\rm de\ Sitter} \right) \right] =
\exp \left[ - \frac{\left( 2 + \sqrt{3}
\right) \pi}{2 \alpha}  \right],
\eeq
by Eq.~\rf{eq-pcr-usual}.
As $\Lambda_{\rm eff}$ decreases further, the pair creation rate
starts to decrease and soon becomes vastly suppressed. For
\beq
0 < \Lambda_{\rm eff} < \Lambda^{\rm C} = \frac{3}{4} \alpha,
\eeq
the lukewarm instanton has the lowest action,
\beq
I_{\rm luke} = - \frac{3\pi}{\Lambda_{\rm eff}} + 
\pi \sqrt{\frac{3}{\alpha \Lambda_{\rm eff}}}.
\eeq
It will therefore dominate the black hole production until the end of
inflation.

If we ask how many charged black holes were produced during the
entire inflationary era, we have to take into account that the density
of the black holes pair created at the early stages of inflation will
be reduced by the subsequent inflationary expansion. As a consequence
most of the charged primordial black holes in our universe were
produced near the end of inflation, as we shall show here.
The number of such black holes per Hubble volume pair created
during one Hubble time when $\phi=\phi_{\rm pc}$ is given by
\beq
\Gamma(\phi_{\rm pc}) = \exp \left[ - \left( I_{\rm luke} 
             - I_{\rm de\ Sitter} \right) \right]
   = \exp \left[ - \sqrt{\frac{3}{\alpha}} \frac{\pi}{m \phi_{\rm pc}}
             \right].
\eeq
At the end of inflation, these black holes have a number density per
Hubble volume of
\beq
d_{\rm end}(\phi_{\rm pc}) =  \Gamma(\phi_{\rm pc}) \,
   \phi_{\rm pc}^3 \exp \left[ - \frac{3}{2} \phi_{\rm pc}^2 \right],
\eeq
where the cubic factor is due to the growth of the Hubble radius and
the exponential factor reflects the inflationary expansion of space.

Since we are dealing with exponential suppressions, practically all
primordial magnetically charged 
black holes in the universe were created at the value of $ \phi_{\rm
pc} $ which makes $ d_{\rm end} $ maximal. This occurs for
\beq
\phi_{\rm pc}^{\rm max} \approx \left( \frac{\pi^2}{3 \, \alpha m^2}
\right)^{1/6},
\eeq
so that the approximate total number per Hubble volume at the
end of inflation is given by
\beq
D_{\rm end} \approx d_{\rm end}\left(\phi_{\rm pc}^{\rm max}\right)
   \approx \left( \frac{\pi^2}{3 \, \alpha m^2} \right)^{1/2}
      \exp \left[ - \left( \frac{9 \pi^2}{16 \, \alpha m^2}
      \right)^{1/3} \right].
\eeq

We take the values of the Hubble radius to be
\beq
H_{\rm end}^{-1} = m^{-1},\ H_{\rm eq}^{-1} = 10^{54},\
H_{\rm now}^{-1} = 10^{59},
\eeq
respectively at the end of inflation, at the time of equal radiation
and matter density, and at the present time. Therefore since the end
of inflation the density has been reduced by a factor of $10^{-91}
m^{-3/2} $ and the Hubble volume has increased by $10^{177} m^3$. We
multiply $ D_{\rm end} $ by these factors to obtain the number of
primordial black holes in the presently observed universe:
\beq
D_{\rm now} \approx 10^{86}
      \left( \frac{\pi^2 m}{3 \, \alpha} \right)^{1/2}
      \exp \left[ - \left( \frac{9 \pi^2}{16 \, \alpha m^2}
      \right)^{1/3} \right].
\eeq
With $ \alpha = 1/137 $ and $ m = 10^{-6} $, the exponent will be on
the order of $ - 10^5 $.  Thus, in ordinary Einstein--Maxwell theory,
it is very unlikely that the observed universe contains even one
magnetic black hole.

However, in theories with a dilaton both the value of the electric
charge and the effective Newton's constant can vary with time. If both
were much higher in the past, the effective value of $ \alpha m^2 $
would not be so small and the present number of magnetic black holes
could be much higher. We are currently working on this question.

\sect{Summary and Conclusions} \label{sec-summary}

Since the quantum pair creation of black holes can be investigated
semi--classically using instanton methods, it has been widely used as
a theoretical laboratory to obtain glimpses at quantum gravity. The
inflationary era is the only time when we can reasonably expect the
effect to have taken place in our own universe.  We chose to work in a
very simple model of chaotic inflation, which allowed us to expose
quite clearly all the important qualitative features of black hole
pair creation.  The inflationary universe was approximated as a
de~Sitter solution with a slowly varying cosmological constant.
Similarly, neutral black holes produced during inflation were
described by a degenerate Schwarzschild--de~Sitter solution. Their
pair creation rate was estimated from the no boundary proposal, by
comparing the probability measures assigned to the two solutions. We
found that Planck size black holes are plentifully produced, but at
later stages in inflation, when the black holes would be larger, the
pair creation rate is exponentially suppressed. This fits in with the
usual instanton prescriptions for pair creation. The tunnelling
proposal, on the other hand, fails to make physically reasonable
predictions. The consideration of black hole pair creation thus lends
support to the no boundary proposal.

We analysed the classical and quantum evolution of neutral primordial
black holes. Classically, the black hole horizon and the cosmological
horizon have the same area and temperature. The two horizons grow
slowly during inflation. We showed that this is due to the classical
flow of scalar field energy across them, according to the First Law of
black hole mechanics. Quantum effects, however, prevent the geometry
from being perfectly degenerate, causing the black hole to be hotter
than the cosmological horizon. As a consequence, practically all
neutral black holes evaporate before the end of inflation.

Finally, we turned to magnetically charged black holes, which can also
be pair created during inflation. Even if they have only one unit of
charge, they cannot evaporate completely and would still exist today.
The pair creation rate is highest during the early stages of
inflation, when the effective cosmological constant is still
relatively large. Black holes created at that time, however, will be
diluted by the inflationary expansion. Most of the charged primordial
black holes were therefore created near the end of the inflationary
era, where they would not be diluted as strongly.  However, in
Einstein--Maxwell theory they are so heavily suppressed that we must
conclude that there are no primordial black holes in the observable
universe. It will therefore be interesting to examine inflationary
models that include a dilaton field.  One would expect to obtain a
much higher number of primordial black holes, which may even allow us
to constrain some of these models.

\section*{Acknowledgements}

We thank Gary Gibbons and Simon Ross for many interesting
discussions. R.\ B.\ gratefully acknowledges financial support from
EPSRC, St John's College and the Studienstiftung des
deutschen Volkes.

\setcounter{section}{0}
\renewcommand{\thesection}{Appendix}
\renewcommand{\theequation}{\Alph{section}.\arabic{equation}}

\sect{}

In this Appendix we show how to calculate the surface gravities of the
two horizons in the Schwarzschild--de~Sitter solution. This
space--time possesses a regular Euclidean section only in the
degenerate (Nariai) case, where the two horizons have the same radius.
Neutral black holes pair created during inflation will therefore start
out nearly degenerate. We present a suitable coordinate transformation
for the nearly degenerate metric, introducing a small parameter
$\eps$, which parametrises the deviation from degeneracy. The surface
gravities and Euclidean action are calculated to second order in
$\eps$, yielding a negative mode in the action. We explain why our
results differ from those obtained in Ref.~\cite{GinPer83}.

The Lorentzian Schwarzschild--de~Sitter solution has the metric
\beq
ds^2 = - U(r) dt^2 + U(r)^{-1} dr^2 + r^2 d\Omega_2^2,
\label{eq-sds-metric}
\eeq
where
\beq
U(r) = 1 - \frac{2 \mu}{r} - \frac{1}{3} \Lambda r^2.
\eeq
For $ 0 < \mu < \frac{1}{3} \Lambda^{-1/2} $, $ U $ has two positive
roots $r_{\rm b}$ and $r_{\rm c}$,
corresponding to the cosmological and the black
hole horizon.
The spacetime admits a timelike Killing vector field
\beq
K = \gamma_t \frac{\partial}{\partial t},
\eeq
where $ \gamma_t $ is a normalisation constant.
The surface gravities $\kappa_{\rm b}$ and $\kappa_{\rm c}$, given by
\beq
\kappa_{\rm b,\:c}
 = \lim_{r \rightarrow r_{\rm b,\:c}} \left[
   \frac{\left( K^a \nabla_a K_b \right) \left( K^c \nabla_c K^b
   \right)}{-K^2} \right]^{1/2},
\label{eq-calc-surface-gravity}
\eeq
depend on the choice of $ \gamma_t $. To obtain the correct value for
the surface gravity, one must normalise the Killing vector in the
right way. In the Schwarzschild case ($ \Lambda = 0 $) the natural
choice is to have $ K^2 = -1 $ at infinity; this corresponds to $
\gamma_t = 1 $ for the standard Schwarzschild metric.  However, in our
case there is no infinity, and it would be a mistake to set $ \gamma_t
= 1 $. Instead one needs to find the radius $ r_{\rm g} $ for which
the orbit of the Killing vector coincides with the geodesic going
through $ r_{\rm g} $ at constant angular variables. This is the
two--sphere at which the effects of the cosmological expansion and the
black hole attraction balance out exactly. An observer at $ r_{\rm g}
$ will need no acceleration to stay there, just like an observer at
infinity in the Schwarzschild case. One must normalise the Killing
vector on this ``geodesic orbit''.  Note that this is a general
prescription which will also give the correct result in the
Schwarzschild limit.  It is straightforward to show that
\beq
r_{\rm g} = \left( \frac{3 \mu}{\Lambda} \right)^{1/3},
\eeq
so that
\beq
\gamma_t = U(r_{\rm g})^{-1/2}
  = \left[ 1 - \left( 9 \Lambda \mu^2 \right)^{1/3} \right]^{-1/2}.
\eeq
Eq.~\rf{eq-calc-surface-gravity} then yields
\beq
\kappa_{\rm b,\:c} = \frac{1}{2 \sqrt{U(r_{\rm g})}}
   \left| \frac{\partial U}{\partial r}
   \right|_{r = r_{\rm b,\:c}}.
\eeq

In order to consider the pair production of black holes, we need to
find a Euclidean instanton which can be analytically continued to the
metric~\rf{eq-sds-metric}. The obvious ansatz is
\beq
ds^2 =  U(r) d\tau^2 + U(r)^{-1} dr^2 + r^2 d\Omega_2^2,
\label{eq-euclidean-sds-metric}
\eeq
where $\tau$ is Euclidean time.
Again one can define a constant $\gamma_{\tau}$ which will normalise
the timelike Killing vector on the geodesic orbit.
In order to avoid a conical singularity at a horizon one needs to
identify $ \tau $ with an appropriate
period $ \tau^{\rm id} $, which is related to
the surface gravity on the horizon by
\beq
\tau^{\rm id} = 2 \pi \gamma_{\tau} / \kappa.
\eeq

Usually only one of the two horizons can be made regular in this way,
since their surface gravities will be different.  They will be equal
only for $ \mu = \frac{1}{3} \Lambda^{-1/2} $, when the two roots of
$U$ coincide. In this degenerate case one can remove both conical
singularities simultaneously and obtains a regular instanton. As was
first pointed out in~\cite{GinPer83}, the fact that $ r_{\rm b} =
r_{\rm c} $ does not mean that the Euclidean region shrinks to zero.
The coordinate system~\rf{eq-euclidean-sds-metric} clearly becomes
inappropriate when $\mu$ approaches its upper limit: the range of $ r
$ becomes arbitrarily narrow while the metric coefficient $ U(r)^{-1}
$ grows without bound. One must therefore perform an appropriate
coordinate transformation. If we write
\beq
9 \mu^2 \Lambda = 1 - 3 \eps^2, \;\; 0 \leq \eps \ll 1,
\eeq
the degenerate case corresponds to $ \eps \rightarrow 0 $. We then
define new time and radial coordinates $ \psi $ and $ \chi $ by
\beq
\tau = \frac{1}{\eps\sqrt{\Lambda}} \left(1
     - \frac{1}{2} \eps^2 \right) \psi; \;\;\;
r    = \frac{1}{\sqrt{\Lambda}} \left[1 + \eps\cch
     - \frac{1}{6} \eps^2 + \frac{4}{9} \eps^3 \cch \right].
\label{eq-transformations}
\eeq
With this choice of $ \psi $ we have
$ \gamma_\psi = \sqrt{\Lambda} $ to second order in $ \eps $, so that
the Killing vector
\beq
K^a = \sqrt{\Lambda} \frac{\partial}{\partial \psi}
\label{eq-eucl-killing-vector}
\eeq
has unit length on the geodesic orbit.  We have chosen the new radial
coordinate $ \chi $ so that $ U $ vanishes to forth order in $ \eps $
for $ \cch = \pm 1 $. This is necessary since $ U $ contains no zero
and first order terms and we intend to calculate all quantities to
second non--trivial order in $ \eps $:
\beq
U(\chi) = \sin^2 \! \chi \; \eps^2 \left[ 1 - \frac{2}{3}\eps\cch +
     \frac{2}{3}\eps^2\csq + \frac{8}{9}\eps^2 \right].
\eeq
Thus the black hole horizon corresponds to $ \chi = \pi $ and the
cosmological horizon to $ \chi = 0 $.

The new metric obtained from the coordinate
transformations~\rf{eq-transformations} is
\beqa
ds^2 & = & \frac{1}{\Lambda} \left( 1 -
                  \frac{2}{3}\eps\cch + \frac{2}{3}\eps^2\csq -
                  \frac{1}{9}\eps^2 \right) \sin^2\!\chi \; d\psi^2
\nonumber \\
     & + & \frac{1}{\Lambda} \left( 1 + \frac{2}{3}\eps\cch -
                  \frac{2}{9}\eps^2\csq \right) d\chi^2
\nonumber \\
     & + & \frac{1}{\Lambda} \left( 1 + 2\eps\cch + \eps^2\csq -
                  \frac{1}{3}\eps^2 \right) d\Omega_2^2.
\label{eq-metric-eps}
\eeqa
In the degenerate case, $ \eps = 0 $, this is the Nariai metric: the
topological product of two round two--spheres, each of radius
$1/\sqrt{\Lambda}$.  There are no conical singularities if the
Euclidean time $ \psi $ is identified with a period $2\pi$. For
general $ \eps $ the two horizons cannot be made regular
simultaneously; it is clear from a geometrical standpoint that $\psi$
must be identified with period
\beq
\psi^{\rm id}_{\rm c,\:b} = 2 \pi \left. \sqrt{g_{\chi\chi}}
   \right|_{\chi=0, \; \pi} \left( \left. \frac{\partial}{\partial
   \chi} \sqrt{g_{\psi\psi}} \right|_{\chi=0, \; \pi} \right)^{-1}
 = 2 \pi \left(1 \pm \frac{2}{3}\eps - \frac{1}{6}\eps^2 \right)
\label{eq-period-eps}
\eeq
to prevent a conical singularity at the cosmological or black hole
horizon, respectively.

One can calculate the surface gravities $\kappa_{\rm c}$ and
$\kappa_{\rm b}$ using the
Euclidean version of Eq.~\rf{eq-calc-surface-gravity} and the Killing
vector~\rf{eq-eucl-killing-vector}; equivalently, one could use the
relation $\kappa = 2 \pi \gamma_{\psi} / \psi^{\rm id}$
to obtain the same result:
\begin{equation}
  \kappa_{\rm c,\:b} = \sqrt{\Lambda} \left(1 \mp \frac{2}{3}\eps +
  \frac{11}{18}\eps^2 \right).
  \label{eq-kappa-eps}
\end{equation}
This equation is useful for the analysis of the radiation energy flux
in a nearly degenerate Lorentzian Schwarzschild--de~Sitter universe,
since each horizon radiates approximately thermally with the
temperature $T = \kappa/2\pi$.

We will now calculate the Euclidean action of the
metric~\rf{eq-metric-eps} and show that it posesses a negative mode in
the direction of decreasing black hole mass.  The total instanton
action is given by
\beq
I = - \frac{\Lambda  {\cal V}}{8 \pi}
    - \frac{A_{\rm c} \delta_{\rm c}}{8 \pi}
    - \frac{A_{\rm b} \delta_{\rm b}}{8 \pi},
\label{eq-action-VA}
\eeq
where ${\cal V}$ is the four-volume of the geometry, $A_{\rm c,\:b}$
are the horizon areas and $\delta_{\rm c,\:b}$ are the conical deficit
angles at the horizons. Of course, all of these quantities depend on
the value we choose for $\psi^{\rm id}$. Obvious options are: to leave
it at $2\pi$ even in the non-degenerate case, thus introducing a
deficit angle on the cosmological horizon and an excess angle at the
black hole horizon, or as an alternative, to make one of the two
horizons regular, thereby causing a larger excess or deficit at the
other horizon. The most interesting of these cases is the one in which
we choose a regular cosmological horizon. In this case, the
metric~\rf{eq-metric-eps} will lie on the interpolation between the
Euclidean Nariai and de~Sitter universes, since the latter has only a
cosmological horizon, which ought to be regular.  The Euclidean
actions for these universes are $-2\pi/\Lambda$ for the Nariai, and
$-3\pi/\Lambda$ for the de~Sitter. Since no
intermediate solution
is known, one would expect the action to decrease
monotonically as one moves away from the Nariai solution. In other
words, this particular perturbation of the Nariai metric should
correspond to a negative mode in the action. Indeed, if we identify
$\psi$ with the period $\psi^{\rm id}_{\rm c}$, the
action in Eq.~\rf{eq-action-VA} turns out as
\beq
I = - \frac{2\pi}{\Lambda} - \frac{17\pi}{9\Lambda}\eps^2 + O(\eps^4).
\eeq
(The same result is obtained for the period $\psi^{\rm id}_{\rm b}$,
while for $ \psi^{\rm id} = 2\pi $ the negative mode is given by $ -
20 \pi \eps^2 / 9 \Lambda $.)

The coordinate transformations, the perturbed Nariai metric and the
negative mode given here differ from Ref.~\cite{GinPer83} for various
reasons. The authors did not ensure that $U=0$ on the horizons, and
the Killing vector wasn't renormalised properly. Also, they identified
Euclidean time with period $2\pi$ even in the non-degenerate case,
which is not appropriate to the physical situation we are trying to
analyse.

\end{document}